# Acoustic Phonon Dispersion Engineering in Bulk Crystals via Incorporation of Dopant Atoms

**Fariborz Kargar[1,2,×], Elias H. Penilla[2,3,×], Ece Aytan[1,2], Jacob S. Lewis[1], Javier E. Garay[2,3*], and Alexander A. Balandin[1,2,*]**

[1]Phonon Optimized Engineered Materials (POEM) Center, Bourns College of Engineering, University of California, Riverside, California 92521 USA

[2]Spins and Heat in Nanoscale Electronic Systems (SHINES) Center, University of California, Riverside, California 92521 USA

[3]Advanced Materials Processing and Synthesis (AMPS) Laboratory, Department of Mechanical and Aerospace Engineering, University of California, San Diego, California 92093 USA

## Abstract

We report results of Brillouin – Mandelstam spectroscopy of transparent $Al_2O_3$ crystals with *Nd* dopants. The ionic radius and atomic mass of *Nd* atoms are distinctively different from those of the host *Al* atoms. Our results show that even a small concentration of *Nd* atoms incorporated into the $Al_2O_3$ samples produces a profound change in the acoustic phonon spectrum. The velocity of the transverse acoustic phonons decreases by ~600 m/s at the *Nd* density of only ~0.1 %. Interestingly, the decrease in the phonon frequency and velocity with the doping concentration is non-monotonic. The obtained results, demonstrating that modification of the acoustic phonon spectrum can be achieved not only by nanostructuring but also by doping have important implications for thermal management as well as thermoelectric and optoelectronic devices.

---

[×] Contributed equally to the work.
[*] Corresponding authors: balandin@ece.ucr.edu and jegaray@eng.ucsd.edu

Acoustic phonons make a dominant contribution to thermal transport in electrical insulators and semiconductors and, at the same time, scatter electrons and holes, limiting the mobility of the charge carriers[1]. Recent years have witnessed a strong increase of interest to the methods of controlling acoustic phonon transport and scattering by tuning the phonon spectrum, *i.e.* dispersion relation, $\omega(q)$ (here $\omega$ is the phonon frequency and $q$ is the phonon wave vector)[2,3]. A possibility of engineering the phonon spectrum provides an additional tuning capability for changing thermal conductivity over the conventional approach, which involves phonon – boundary scattering[4–7]. It also allows one to affect the way phonons interact with electrons and light. Until now, the *phonon engineering* approach has been associated with the nanostructured materials, where the phonon dispersion undergoes modification due to the periodic or stationary boundary conditions imposed in addition to the periodicity of the atomic crystal structure[4–11]. In this method, nanometer scale dimensions are essential in order to reveal the wave nature of the phonons and induce modification in their spectrum. The average "gray" phonon mean free path (MFP), $\Lambda$, determined from the expression $\Lambda = 3K/C_p \upsilon$ is on the order of 10 nm – 100 nm at room temperature (RT) for many materials[12] (here $K$ is the phonon thermal conductivity, $C_p$ is the specific heat and $\upsilon$ is the phonon group velocity). The latter estimates explain the need for structuring the material and nanometer scale. A strong modification of the acoustic phonon dispersion has been demonstrated in numerous periodic phononic crystals[8–11] and even in individual semiconductor nanowires[13].

In this Letter, we describe a drastically different approach for changing the acoustic phonon spectrum of the materials, which does not rely on nanostructuring. Our results show that one can engineer the phonon spectrum in bulk crystals via doping with atoms that have a substantially different mass and size from those of the host atoms. The importance of the obtained data goes beyond the developments of the traditional phonon engineering approaches. A noticeable decrease in the phonon velocity with even extremely small concentration of foreign atoms (~0.1 %) means that the theoretical description of thermal conductivity should be adjusted accordingly. Indeed, in the Callaway – Klemens method, and other Boltzmann transport equation approaches, the effects of doping are accounted for by introduction of the phonon – point defect scattering term proportional to the defect density concentration, or fraction of the foreign atoms[14,15]. This treatment always assumes that the velocity of acoustic phonons itself does not change with the doping. Our results demonstrate that the latter is not the case.

Commercially available $Al_2O_3$ (99.99% purity, Taimei Chemicals, Japan) was processed using current activated pressure assisted densification (CAPAD)[16] using a similar method described in [17]. Briefly, the as received (un-doped) and $Nd_2O_3$ doped (99.99% purity, Alfa Aesar, USA) powders were mixed to achieve a doping level (Nd3+: Al3+) of 0.1-0.5 at.%. using water and low energy ball milling, followed by planetary ball milling before densification in CAPAD. X-ray diffraction was taken on a Phillips instrument (X'Pert Model: DY1145), operating in the point source mode, 45 kV potential and 40 mA current was taken on a with a 0.002 step size and 4 s integration time. The experimental data were fitted using Gaussian profiles, with the K-$\alpha_2$ contribution removed numerically.

Figure 1a is an SEM (Zeiss Sigma 500) micrograph of a fracture surface of the polycrystalline 0.25 Nd: $Al_2O_3$ revealing ~300 nm equiaxed grain structure. The inset in Figure 1a shows a picture of a representative sample on top of printed text. Our samples are transparent because samples with <0.4 at% *Nd* do not have second phases and they have low *Nd* segregation to grain boundaries and grain triple points. Other phases and segregation or both would scatter light because of refractive index mismatch, significantly reducing transparency. Figure 1 b are XRD patterns near the (113) plane of the $Al_2O_3$ and *Nd*: $Al_2O_3$ samples. Also plotted is an ICSD standard (#63647) for comparison. The pure $Al_2O_3$ shows a peak at the same 2θ location as the standard confirming the crystal structure, while the doped samples show a clear peak shift to lower angles, indicating an expansion of the lattice which is expected from the *Nd* doping since the radius of *Nd* atoms is significantly larger than *Al*.

[Figure 1: Samples]

The structure of the samples was also confirmed by Raman spectroscopy (Renishaw InVia). The measurements have been conducted in the backscattering configuration under 488 nm excitation. Figure 2 shows the Raman spectra of pristine $Al_2O_3$ and samples with 0.1 % and 0.25% of *Nd* dopants. We resolved seven distinct optical phonon peaks at 379.0, 418.6, 430.7, 447.1, 576.9, 645.8 and 751.0 $cm^{-1}$. The Raman spectrum is in excellent agreement with the previously reported studies[18–20]. The peaks at 418.6 $cm^{-1}$ and 645.8 $cm^{-1}$ belong to $A_{1g}$ vibrational modes while the others are associated with $E_g$ vibrational bands. The relative intensity of these phonon modes varies in accordance with the crystallographic directions. It is important to note that all optical phonon modes, with exception of one at 430.7 $cm^{-1}$, show some softening with introduction of *Nd* doping atoms. This is in line with the XRD measurements, which indicated the shift of the ~43.4$^o$ peak to smaller angles, suggesting some distortion in the lattice due to *Nd* incorporation. The implications of this observation from the Raman and XRD studies will be discussed below.

[Figure 2: Raman]

The Brillouin-Mandelstam light scattering spectroscopy (BMS) allows one to directly probe the acoustic phonon frequencies close to the Brillouin zone (BZ) center[21–23]. We conducted BMS studies in the backscattering configuration. In this geometry, the *p*-polarized (the electric field direction of the light parallel to the scattering plane) laser light is focused on the sample by a lens with a numerical aperture of 1.4. The scattered light was collected using the same lens and directed to the high contrast six pass tandem Fabry-Perot interferometer. The spectra were excited with the solid-state diode-pumped laser (Coherent) operating at $\lambda$=532 nm[24,25]. In all experiments, the power on the sample was adjusted to be 70 mW in order to avoid self-heating effects. The incident angle of the laser light with respect to the normal to the sample was fixed at 30$^o$. The in-plane rotation did not affect the BMS results due to the polycrystalline structure of the samples. For each sample, the BMS measurements were repeated several times (>5) by focusing light on different spots, in order to exclude a possibility of material parameters variations over the sample volume. In transparent bulk $Al_2O_3$ samples the volumetric elasto-optic effect is the dominant light scattering mechanism by acoustic phonons[21–23]. For this reason, interpretation of BMS data requires a knowledge of the refractive index, $n$, of the material. The refractive index was measured by the "prism coupling" method (Metricon)[26]. The measurements revealed $n$=1.767 (at $\lambda$=532 nm) in all directions confirming the optically isotropic nature of all the samples. Since the concentration of the $Nd$ dopants is very low (<0.5 wt%), the Maxwell – Garnett approximation, $n = \left( (1-\phi)n_m^2 + \phi n_o^2 \right)^{1/2}$ ($n_m$ and $n_o$ are the refractive indices of $Al_2O_3$ and $Nd$, respectively) described the trivial variation in $n$ with high accuracy. The probe phonon wave vector, $q = 4\pi n / \lambda$, for the elasto-optic scattering mechanism was determined to be $q$=0.0417 nm$^{-1}$ in our samples.

Figure 3 shows BMS data for $Al_2O_3$ with and without $Nd$ dopants. One can see in Figure 3 (a) that three peaks corresponding to one longitudinal (LA) and two transverse (TA$_1$ and TA$_2$) acoustic phonon polarization branches are resolved. The peaks have been fitted and deconvoluted, when required, with individual Lorentzian functions (green curves). The red curve in Figure 3 (a) is the cumulative fitting to the experimental data. For alumina, the LA and two TA peaks were found at 73.0, 45.6 and 41.3 GHz, respectively. Adding the $Nd$ dopants resulted in an unexpected large decrease in the frequency of all three acoustic phonon peaks. Figure 3 (b) shows the evolution of the LA and TA phonon polarization branches in $Al_2O_3$ as the concentration of $Nd$ dopants increases from 0 to 0.50 %. Incorporation of $Nd$ dopants results in pronounced decrease of the intensity of the phonon peaks and reduction in their frequency. The decrease in the intensity, which was also observed in Raman spectra (see Figure 2) was attributed to the increase in the light absorption (especially at $Nd$ concentration ≥0.4), and corresponding decrease in the interaction volume for BMS. While the intense LA peak was persistent at all $Nd$ concentrations, the TA peaks could not be resolved in the sample with 0.50% of $Nd$ atoms.

[Figure 3: BMS Spectra]

Figure 4 (a-c) shows the variation of the peak position of LA and TA phonons with $Nd$ concentration. The frequency mean value and standard deviation have been determined from measurements on different spots of each samples. The results show a surprising sharp decrease in the acoustic phonon frequency with the concentration of $Nd$ as low as 0.1%. This was observed consistently for each phonon polarization branch. Adding more dopants, up to 0.4%, leads to much weaker decrease in the phonon frequency. The frequency of LA phonon starts to decrease faster again as $Nd$ concentration reaches 0.5%. Overall, the frequencies of TA branches are affected stronger than that of LA branch. While the frequency difference of LA phonons for pristine alumina and 0.4% $Nd$: $Al_2O_3$ is ~1 GHz, it exceeds ~4.5 GHz for TA modes. BMS probes the phonons with wave vector close to BZ center. Since in this region, the dispersion of the acoustic phonons is linear and $\omega(q=0)=0$, one can determine the phonon group velocity (sound velocity) knowing the frequency at one value of the phonons wave vector. Table I reports the group velocity of different branches as a function of $Nd$ concentration.

[Figure 4]

[Table I]

Elasticity theory relates the phonon velocity to material properties as $\upsilon = (E/\rho)^{1/2}$ where $E$ is the Young's modulus and $\rho$ is the mass density. Since the concentration of $Nd$ is extremely small one would expect equally small changes due to $\rho$ variation. The probable scenario is that the introduction of $Nd$ dopants changes the elastic properties of the lattice. The atomic mass difference, although large between $Al$ and $Nd$ (factor of ×5) is unlikely to change the frequency of vibrations, again due to the small concentration of dopants. A possible mechanism can be related to the lattice distortion created by larger $Nd$ atoms (factor of ×2 bigger than $Al$), which is accompanied by increased atomic plane separation, in line with Raman and XRD data. This can also account for the observed abrupt decrease at the smallest concentration of $Nd$ (0.1%) followed by a much weaker dependence at higher $Nd$ concentrations: adding a few more atoms would not substantially increase the plane separation. Complete understanding of the mechanism of the phonon frequency change requires detailed microscopic studies and ab initio theory, which goes beyond the scope of this work.

It is interesting to assess implications of the phonon velocity reduction for thermal transport and phonon interaction with other elemental excitations. The phonon thermal conductivity can be written as $K = (1/3)C\upsilon\Lambda = (1/3)C\upsilon^2\tau$, where the phonon life-time, $\tau$, can be expressed through the phonon scattering rates in two main relaxation mechanisms: anharmonic phonon Umklapp scattering and phonon – point defect scattering so that $\tau^{-1} = \tau_U^{-1} + \tau_P^{-1}$, where $\tau_U^{-1}$ and $\tau_P^{-1}$ are scattering rates in the Umklapp and point defect processes, respectively. The phonon scattering rate on point defects, $1/\tau_P \propto V_0(\omega^4/\upsilon^3)\Gamma$, where $V_0$ is the volume per one atom in the crystal lattice, $\omega$ is the phonon frequency and $\Gamma$ is the strength of the phonon - point defect scattering, which depends on the fraction of the foreign atoms. In the perturbation theory $\Gamma$ is written as[14,15]

$$\Gamma = \sum_i f_i \left[ \left(1 - M_i/\overline{M}\right)^2 + \varepsilon\left(\gamma\left(1 - R_i/\overline{R}\right)\right)^2 \right] \qquad (1)$$

Here $f_i$ is the fractional concentration of the substitutional foreign atoms, $M_i$ is the mass of the $i$th substitutional atom, $\overline{M} = \sum_i f_i M_i$ is the average atomic mass, $R_i$ is the Pauling ionic radius of the $i$th foreign atom, $\overline{R} = \sum_i f_i R_i$ is the average radius, $\gamma$ is the Grüneisen parameter, and $\varepsilon$ is a phenomenological parameter. One can see from these formulas that the effect of dopants on thermal conductivity is accounted via $\tau_P^{-1}$ term: it is proportional to the concentration and growth with the increasing difference in the atomic radius and mass between the host the foreign atoms. Conventional theory does not consider any variation in the phonon velocity and assumes that $\upsilon$ remains unchanged after the doping. Our results show that in alumina (and presumably in other rather common materials) and at very small concentration of dopants the assumption of the constant phonon velocity is not strictly valid. The effect of dopant introduction can be amplified via the velocity change and affect the thermal conductivity values, particularly at low temperature where Umklapp scattering is minimal. The effects of the phonon velocity reduction are not limited to heat conduction. They can also reveal themselves in electron – acoustic phonon scattering, electron – photon interaction, which involve phonons, and electron – hole non-radiative recombination. Examples of the latter include phonon-assisted non-radiative recombination in the Auger processes, where electron – acoustic phonon coupling is inversely proportional to the phonon group velocity[27,28].

In summary, we observed that the velocity of acoustic phonons can be changed by even a small concentration of dopants with distinctively different atomic size and mass. The obtained results, demonstrating a possibility of phonon engineering in bulk crystals have important implications for thermal management as well as thermoelectric and optoelectronic devices.

*Acknowledgements*

The material synthesis and characterization work was supported as part of the Spins and Heat in Nanoscale Electronic Systems (SHINES), an Energy Frontier Research Center funded by the U.S. Department of Energy, Office of Science, Basic Energy Sciences (BES) under Award # SC0012670. AAB also acknowledges the support of DARPA project W911NF18-1-0041 Phonon Engineered Materials for Fine-Tuning the G-R Center and Auger Recombination for the work related to the development of the phonon engineering concept. JEG also acknowledges support from High Energy Laser - Joint Technology Office (HEL-JTO) administered by the Army Research Office for development of over-equilibrium doped alumina.

**FIGURE CAPTIONS**

**Figure 1: (a)** SEM micrograph of a fracture surface of the polycrystalline 0.25 at% *Nd*: $Al_2O_3$ revealing ~300 nm equiaxed grain structure. The scale bar is 200 nm. The inset shows a picture of a representative sample on top of printed text, revealing optical transparency **(b)** XRD patterns near the (113) plane of the $Al_2O_3$ and *Nd*: $Al_2O_3$ samples. Also plotted is an ICSD standard (#63647) for comparison.

**Figure 2:** Raman spectrum of the $Al_2O_3$, 0.10 at% *Nd:* $Al_2O_3$ and 0.25 at% *Nd*: $Al_2O_3$. As the amount of *Nd* density increases, the Raman peaks except the one at ~430 $cm^{-1}$ shift to the lower wave numbers. The relative intensity as well as peak position of the Raman peaks confirm the composition, quality and polycrystalline structure of the samples.

**Figure 3:** (a) Brillouin-Mandelstam scattering spectra for the pure $Al_2O_3$ and 0.35 at% *Nd:* $Al_2O_3$. The experimental data (black curve) has been fitted using individual (green curve) and cumulative (red curve) Lorentzian fittings. The regular longitudinal (LA) and transverse (TA) acoustic phonons are present in both spectra (b) Evolution of the spectrum with increasing the *Nd* doping level. Note the decrease in frequency of LA and TA phonons of pure $Al_2O_3$ with increasing the *Nd* density to 0.1% and more.

**Figure 4:** Peak position of LA and TA phonon polarization branches in Brillouin spectra versus *Nd* density. The frequency of LA and both TA phonon branches decreases with increasing *Nd* concentration non-monotonically. Note that even the smallest 0.1 at% concentration of *Nd* results in a noticeable decrease of the phonon frequency, and, correspondingly, group velocity.

**Table I:** Phonon group velocity in $Al_2O_3$ for different $Nd$ doping concentration

| Sample | TA Group Velocity (m/s) | TA Group Velocity (m/s) | LA Group Velocity (m/s) |
|---|---|---|---|
| $Al_2O_3$ | 6269.0 | 6864.8 | 11012.7 |
| 0.10 at% $Nd$:$Al_2O_3$ | 5670.0 | 6311.7 | 10845.6 |
| 0.25 at% $Nd$:$Al_2O_3$ | 5660.8 | 6320.5 | 10862.8 |
| 0.35 at% $Nd$:$Al_2O_3$ | 5583.1 | 6247.9 | 10858.3 |
| 0.40 at% $Nd$:$Al_2O_3$ | 5539.8 | 6178.1 | 10861.3 |
| 0.50 at% $Nd$:$Al_2O_3$ | - | - | 10802.6 |

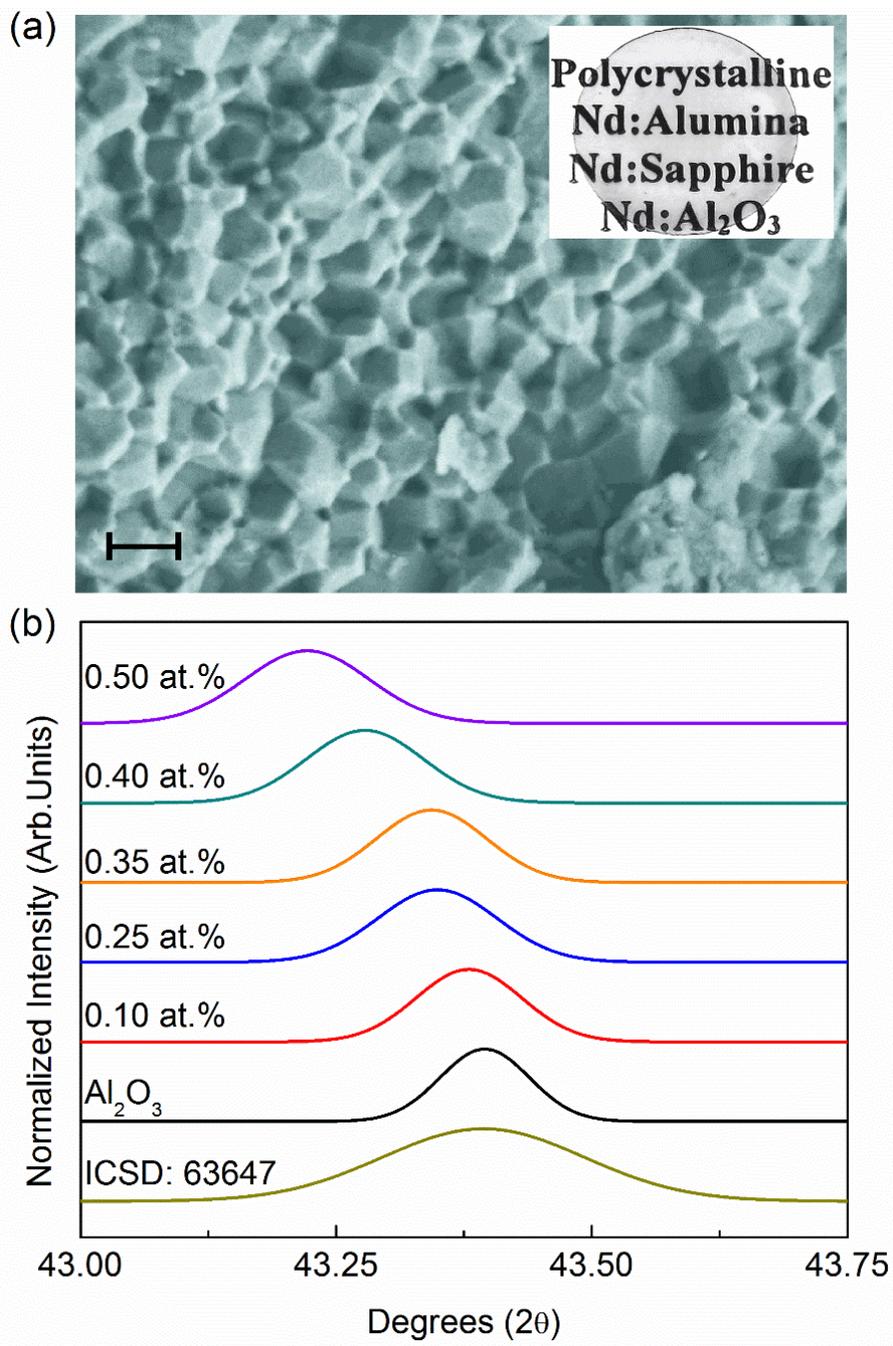

[Figure 1]

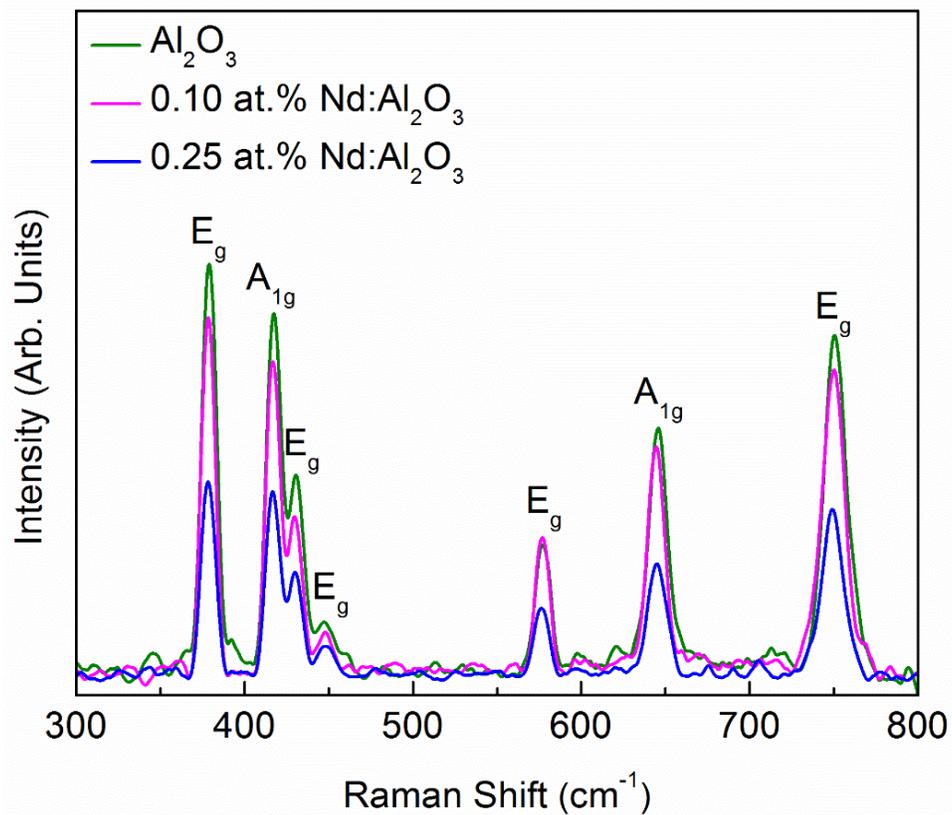

[Figure 2]

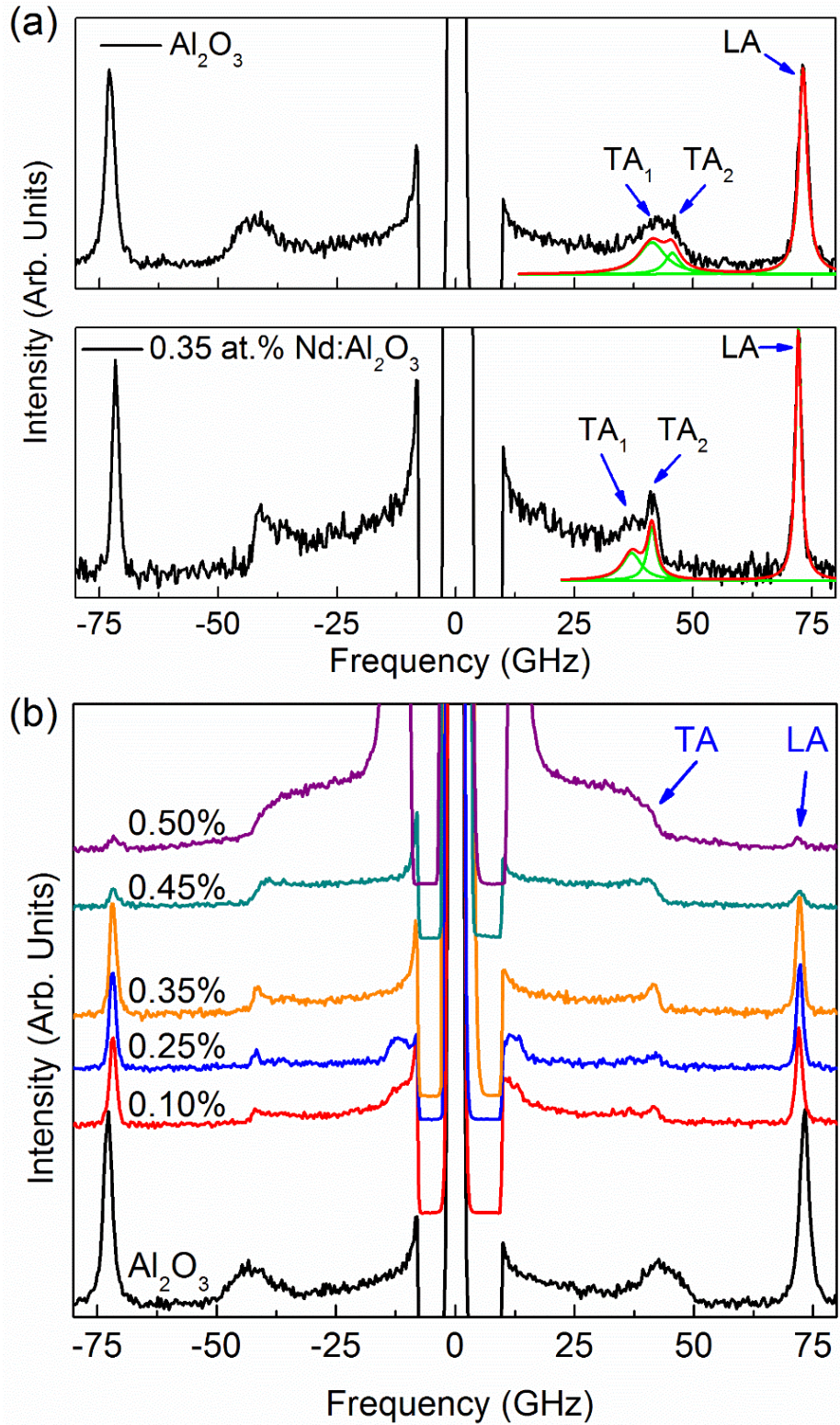

[Figure 3]

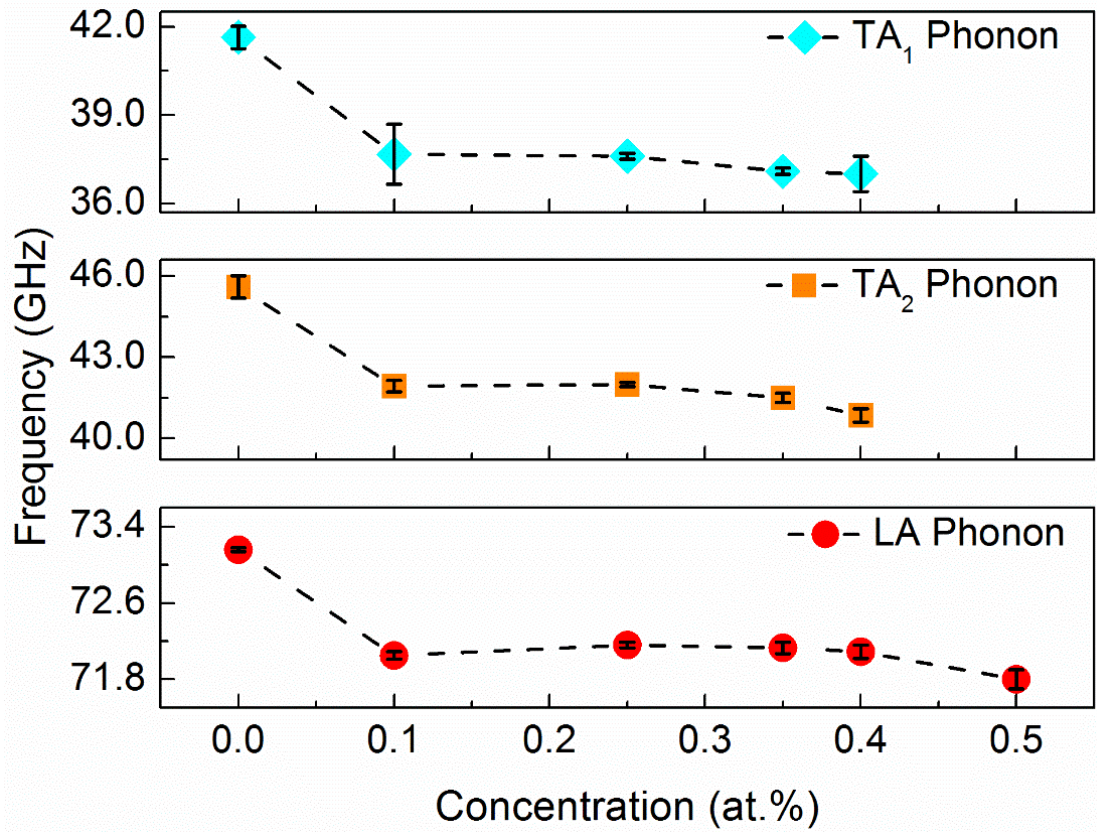

[Figure 4]